\pdfoutput=1
\documentclass[12pt,a4paper]{article}

\usepackage{ifthen} 
\newboolean{pdflatex}
\setboolean{pdflatex}{true} 

\newboolean{articletitles}
\setboolean{articletitles}{true} 

\newboolean{uprightparticles}
\setboolean{uprightparticles}{false} 

\newboolean{inbibliography}
\setboolean{inbibliography}{false} 

\usepackage{microtype}
\usepackage{lineno}  
\usepackage{xspace} 
\usepackage{caption} 

\usepackage{graphicx}  
\usepackage{color}
\usepackage{colortbl}
\graphicspath{{./figs/}} 

\usepackage{amsmath} 
\usepackage{amssymb}
\usepackage{amsfonts}
\usepackage{upgreek} 

\newcommand*\patchAmsMathEnvironmentForLineno[1]{%
\expandafter\let\csname old#1\expandafter\endcsname\csname #1\endcsname
\expandafter\let\csname oldend#1\expandafter\endcsname\csname
end#1\endcsname
 \renewenvironment{#1}%
   {\linenomath\csname old#1\endcsname}%
   {\csname oldend#1\endcsname\endlinenomath}%
}
\newcommand*\patchBothAmsMathEnvironmentsForLineno[1]{%
  \patchAmsMathEnvironmentForLineno{#1}%
  \patchAmsMathEnvironmentForLineno{#1*}%
}
\AtBeginDocument{%
\patchBothAmsMathEnvironmentsForLineno{equation}%
\patchBothAmsMathEnvironmentsForLineno{align}%
\patchBothAmsMathEnvironmentsForLineno{flalign}%
\patchBothAmsMathEnvironmentsForLineno{alignat}%
\patchBothAmsMathEnvironmentsForLineno{gather}%
\patchBothAmsMathEnvironmentsForLineno{multline}%
}

\usepackage{hyperref}    
\usepackage[all]{hypcap} 

\def\lhcb {\mbox{LHCb}\xspace}

\ifthenelse{\boolean{uprightparticles}}%
{

 \def\PDelta      {\ensuremath{\Delta}\xspace}                 
 \def\PXi      {\ensuremath{\Xi}\xspace}                 
 \def\PLambda      {\ensuremath{\Lambda}\xspace}                 
 \def\PSigma      {\ensuremath{\Sigma}\xspace}                 
 \def\POmega      {\ensuremath{\Omega}\xspace}                 
 \def\PUpsilon      {\ensuremath{\Upsilon}\xspace}                 
 
 \def\PB      {\ensuremath{\mathrm{B}}\xspace}                 
                  
 \def\PD      {\ensuremath{\mathrm{D}}\xspace}

 \def\PK      {\ensuremath{\mathrm{K}}\xspace}

 \def\Pb      {\ensuremath{\mathrm{b}}\xspace}                 
 \def\Pc      {\ensuremath{\mathrm{c}}\xspace}

 \def\Pi      {\ensuremath{\mathrm{i}}\xspace}

 \def\Ps      {\ensuremath{\mathrm{s}}\xspace}

}
{

 \mathchardef\PDelta="7101
 \mathchardef\PXi="7104
 \mathchardef\PLambda="7103
 \mathchardef\PSigma="7106
 \mathchardef\POmega="710A
 \mathchardef\PUpsilon="7107
                  
 \def\PB      {\ensuremath{B}\xspace}                 
                  
 \def\PD      {\ensuremath{D}\xspace}

 \def\PK      {\ensuremath{K}\xspace}

 \def\Pb      {\ensuremath{b}\xspace}                 
 \def\Pc      {\ensuremath{c}\xspace}

 \def\Pi      {\ensuremath{i}\xspace}

 \def\Ps      {\ensuremath{s}\xspace}

}

\def\squark    {\ensuremath{\Ps}\xspace}

\def\cquark    {\ensuremath{\Pc}\xspace}

\def\bquark    {\ensuremath{\Pb}\xspace}

  \def\Kbar  {\kern 0.2em\overline{\kern -0.2em \PK}{}\xspace}

  \def\Dbar    {\kern 0.2em\overline{\kern -0.2em \PD}{}\xspace}
\def\D       {\ensuremath{\PD}\xspace}

\def\B       {\ensuremath{\PB}\xspace}
\def\Bbar    {\ensuremath{\kern 0.18em\overline{\kern -0.18em \PB}{}}\xspace}

\def\Bd      {\ensuremath{\B^0}\xspace}
\def\Bs      {\ensuremath{\B^0_\squark}\xspace}

  \def\Y#1S{\ensuremath{\PUpsilon{(#1S)}}\xspace}

\def\Lbar {\ensuremath{\kern 0.1em\overline{\kern -0.1em\PLambda}}\xspace}

\def\Lb      {\ensuremath{\Lz^0_\bquark}\xspace}

\def\Lc      {\ensuremath{\Lz^+_\cquark}\xspace}

\def\to                 {\ensuremath{\rightarrow}\xspace}

\def\AT#1     {\ensuremath{A_{\mathrm{T}}^{#1}}\xspace}           

\def\C#1      {\ensuremath{\mathcal{C}_{#1}}\xspace}                       
\def\Cp#1     {\ensuremath{\mathcal{C}_{#1}^{'}}\xspace}                    
\def\Ceff#1   {\ensuremath{\mathcal{C}_{#1}^{\mathrm{(eff)}}}\xspace}        
\def\Cpeff#1  {\ensuremath{\mathcal{C}_{#1}^{'\mathrm{(eff)}}}\xspace}       
\def\Ope#1    {\ensuremath{\mathcal{O}_{#1}}\xspace}                       
\def\Opep#1   {\ensuremath{\mathcal{O}_{#1}^{'}}\xspace}                    

\newcommand{\tev}{\ifthenelse{\boolean{inbibliography}}{\ensuremath{~T\kern -0.05em eV}\xspace}{\ensuremath{\mathrm{\,Te\kern -0.1em V}}\xspace}}
\newcommand{\gev}{\ensuremath{\mathrm{\,Ge\kern -0.1em V}}\xspace}
\newcommand{\mev}{\ensuremath{\mathrm{\,Me\kern -0.1em V}}\xspace}
\newcommand{\kev}{\ensuremath{\mathrm{\,ke\kern -0.1em V}}\xspace}
\newcommand{\ev}{\ensuremath{\mathrm{\,e\kern -0.1em V}}\xspace}
\newcommand{\gevc}{\ensuremath{{\mathrm{\,Ge\kern -0.1em V\!/}c}}\xspace}
\newcommand{\mevc}{\ensuremath{{\mathrm{\,Me\kern -0.1em V\!/}c}}\xspace}
\newcommand{\gevcc}{\ensuremath{{\mathrm{\,Ge\kern -0.1em V\!/}c^2}}\xspace}
\newcommand{\gevgevcccc}{\ensuremath{{\mathrm{\,Ge\kern -0.1em V^2\!/}c^4}}\xspace}
\newcommand{\mevcc}{\ensuremath{{\mathrm{\,Me\kern -0.1em V\!/}c^2}}\xspace}

\def\mum  {\ensuremath{{\,\upmu\rm m}}\xspace}

\def\invfb   {\ensuremath{\mbox{\,fb}^{-1}}\xspace}

\newcommand{\stat}{\ensuremath{\mathrm{\,(stat)}}\xspace}
\newcommand{\syst}{\ensuremath{\mathrm{\,(syst)}}\xspace}

\def\gsim{{~\raise.15em\hbox{$>$}\kern-.85em
          \lower.35em\hbox{$\sim$}~}\xspace}
\def\lsim{{~\raise.15em\hbox{$<$}\kern-.85em
          \lower.35em\hbox{$\sim$}~}\xspace}

\def\pt         {\mbox{$p_{\rm T}$}\xspace}

\def\evtgen     {\mbox{\textsc{EvtGen}}\xspace}

\def\gauss      {\mbox{\textsc{Gauss}}\xspace}
\def\geant      {\mbox{\textsc{Geant4}}\xspace}

\def\photos     {\mbox{\textsc{Photos}}\xspace}

\def\pythia     {\mbox{\textsc{Pythia}}\xspace}

\def\tell1  {TELL1\xspace}
\def\ukl1   {UKL1\xspace}

\usepackage{cite} 
\usepackage{mciteplus}

\begin{document}

\renewcommand{\thefootnote}{\fnsymbol{footnote}}

\def \LbLcDs {\ensuremath{{\it \Lambda}_b^0 \to {\it \Lambda}_c^+ D^-_s}\xspace}
\def \LbLcD {\ensuremath{{\it \Lambda}_b^0 \to {\it \Lambda}_c^+ D^-}\xspace}
\def \BdDDs {\ensuremath{{\kern 0.2em}\overline{\kern -0.2em B}{}^0 \to D^+ D^-_s}\xspace}
\def \BsDDs {\ensuremath{B_s^0 \to D^+ D^-_s}\xspace}
\def \BdsDDs {\ensuremath{B_{(s)}^0 \to D^+ D^-_s}\xspace}
\def \BdsLcLc {\ensuremath{ B^0_{(s)} \to {\it \Lambda}_c^+ {\it \Lambda}_c^-}\xspace}
\def \BdLcLc {\ensuremath{{\kern 0.2em}\overline{\kern -0.2em B}{}^0 \to {\it \Lambda}_c^+ {\it \Lambda}_c^-}\xspace}
\def \BsLcLc {\ensuremath{ B^0_s \to {\it \Lambda}_c^+ {\it \Lambda}_c^-}\xspace}
\def \Lb {\ensuremath{{\it \Lambda}_b^0}\xspace}
\def \Lc {\ensuremath{{\it \Lambda}_c^+}\xspace}
\def \LbLcpi {\ensuremath{{\it \Lambda}_b^0 \to {\it \Lambda}_c^+ \pi^-}\xspace}
\def \BdDpi {\ensuremath{{\kern 0.2em}\overline{\kern -0.2em B}{}^0 \to D^+ \pi^-}\xspace}
\def \BsDspi {\ensuremath{{\kern 0.2em}\overline{\kern -0.2em B}{}^0_s \to D^+_s \pi^-}\xspace}
\def \Bd {\ensuremath{{\kern 0.2em}\overline{\kern -0.2em B}{}^0}\xspace}
 
\setcounter{footnote}{1}


\begin{titlepage}
\pagenumbering{roman}

\vspace*{-1.5cm}
\centerline{\large EUROPEAN ORGANIZATION FOR NUCLEAR RESEARCH (CERN)}
\vspace*{1.5cm}
\hspace*{-0.5cm}
\begin{tabular*}{\linewidth}{lc@{\extracolsep{\fill}}r}
\ifthenelse{\boolean{pdflatex}}
{\vspace*{-2.7cm}\mbox{\!\!\!\includegraphics[width=.14\textwidth]{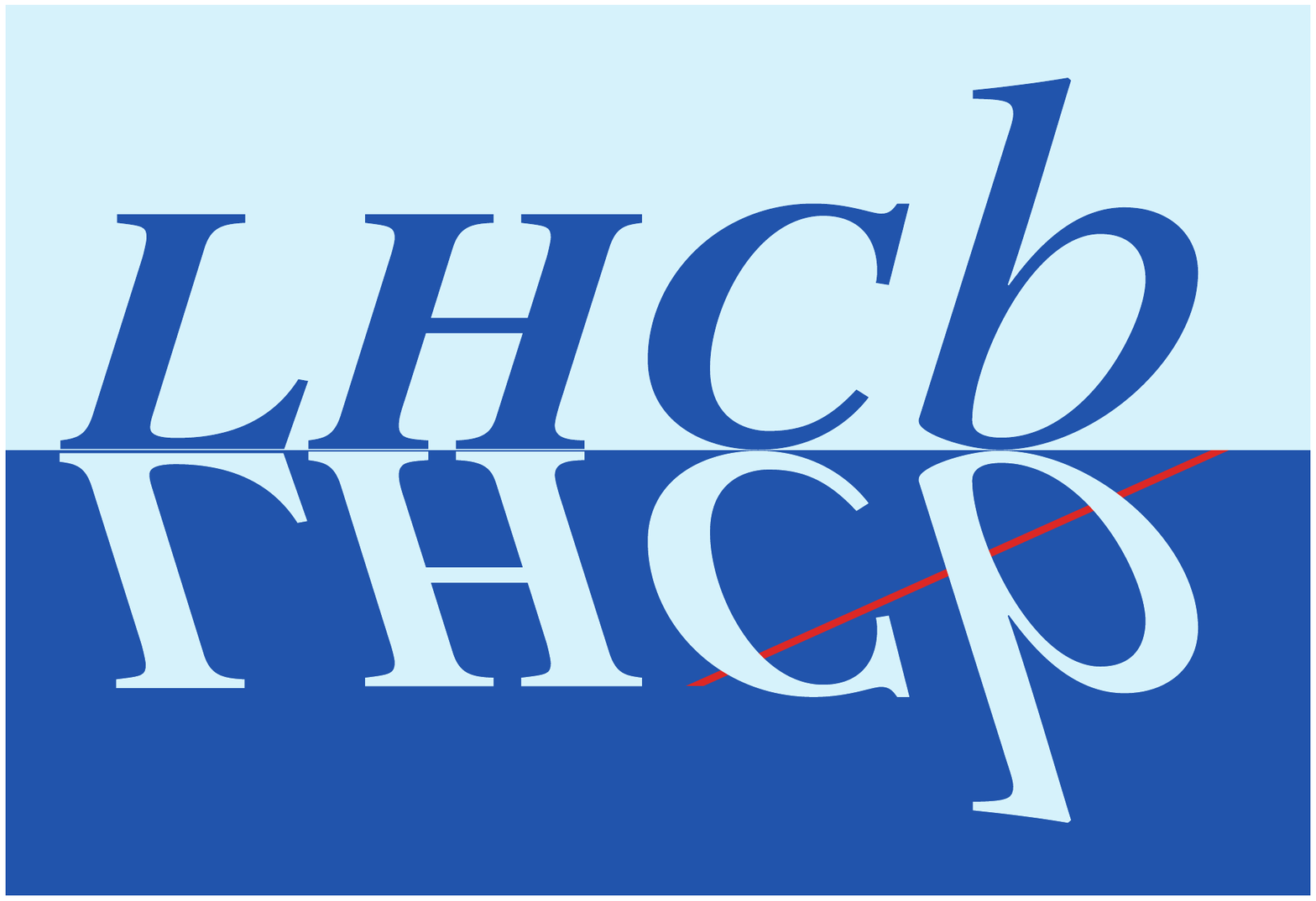}} & &}%
{\vspace*{-1.2cm}\mbox{\!\!\!\includegraphics[width=.12\textwidth]{lhcb-logo.eps}} & &}%
\\
 & & CERN-PH-EP-2014-033 \\  
 & & LHCb-PAPER-2014-002 \\  
 & & March,13 2014 \\ 
 & & \\
\end{tabular*}

\vspace*{0.5cm}

{\bf\boldmath\huge
\begin{center}
Study of beauty hadron decays into pairs of charm hadrons
\end{center}
}

\vspace*{0.3cm}

\begin{center}
The LHCb collaboration\footnote{Authors are listed on the following pages.}
\end{center}

\vspace*{0.2cm}

\begin{abstract}
  \noindent
   First observations of the decays $\Lb \to \Lc D_{(s)}^-$ are reported using data corresponding to an integrated luminosity of 
3\invfb collected at 7 and 8\tev center-of-mass energy in proton-proton collisions with the  \lhcb detector.  In addition, the most precise measurement of the branching fraction $\mathcal{B}(\BsDDs)$ is made and a search is performed for the decays \BdsLcLc.  The results obtained are 
  \begin{eqnarray*}
\mathcal{B}(\LbLcD)/\mathcal{B}(\LbLcDs) &=& 0.042 \pm 0.003\stat \pm 0.003\syst,\\
\left[\frac{\mathcal{B}(\LbLcDs)}{\mathcal{B}(\BdDDs)}\right]\big/\left[\frac{\mathcal{B}(\LbLcpi)}{\mathcal{B}(\BdDpi)}\right] &=& 0.96 \pm 0.02\stat \pm 0.06\syst,\\
\mathcal{B}(\BsDDs)/\mathcal{B}(\BdDDs) &=& 0.038\pm0.004\stat\pm0.003\syst,\\
\mathcal{B}(\BdLcLc)/\mathcal{B}(\BdDDs) & < & 0.0022\; [95\% \; {\rm C.L.}],\\
\mathcal{B}(\BsLcLc)/\mathcal{B}(\BsDDs) & < & 0.30\; [95\% \; {\rm C.L.}].
  \end{eqnarray*}
Measurement of the mass of the \Lb baryon relative to the \Bd meson gives ${M(\Lb) -M(\Bd) = 339.72\pm 0.24\stat \pm 0.18\syst}$\mevcc.  This result provides the most precise measurement of the mass of the \Lb baryon to date.
\end{abstract}

\vspace*{0.2cm}

\begin{center}
  Submitted to Physical Review Letters.  
\end{center}

\vspace{\fill}

{\footnotesize 
\centerline{\copyright~CERN on behalf of the \lhcb collaboration, license \href{http://creativecommons.org/licenses/by/3.0/}{CC-BY-3.0}.}}
\vspace*{2mm}

\end{titlepage}

\newpage
\setcounter{page}{2}
\mbox{~}
\newpage


\centerline{\large\bf LHCb collaboration}
\begin{flushleft}
\small
R.~Aaij$^{41}$, 
A.~Abba$^{21,u}$, 
B.~Adeva$^{37}$, 
M.~Adinolfi$^{46}$, 
A.~Affolder$^{52}$, 
Z.~Ajaltouni$^{5}$, 
J.~Albrecht$^{9}$, 
F.~Alessio$^{38}$, 
M.~Alexander$^{51}$, 
S.~Ali$^{41}$
% ...

{\footnotesize \it
% ...
$ ^{5}$Clermont Universit\'{e}, Universit\'{e} Blaise Pascal, CNRS/IN2P3, LPC, Clermont-Ferrand, France\\
% ...
$ ^{9}$Fakult\"{a}t Physik, Technische Universit\"{a}t Dortmund, Dortmund, Germany\\
% ...
$ ^{21}$Sezione INFN di Milano, Milano, Italy\\
% ...
$ ^{37}$Universidad de Santiago de Compostela, Santiago de Compostela, Spain\\
$ ^{38}$European Organization for Nuclear Research (CERN), Geneva, Switzerland\\
% ...
$ ^{41}$Nikhef National Institute for Subatomic Physics, Amsterdam, The Netherlands\\
% ...
$ ^{46}$H.H. Wills Physics Laboratory, University of Bristol, Bristol, United Kingdom\\
% ...
$ ^{51}$School of Physics and Astronomy, University of Glasgow, Glasgow, United Kingdom\\
$ ^{52}$Oliver Lodge Laboratory, University of Liverpool, Liverpool, United Kingdom\\
% ...
$ ^{u}$Politecnico di Milano, Milano, Italy\\
}
\end{flushleft}

\cleardoublepage

\renewcommand{\thefootnote}{\arabic{footnote}}
\setcounter{footnote}{0}

\pagestyle{plain} 
\setcounter{page}{1}
\pagenumbering{arabic}

Hadrons are systems of quarks bound by the strong interaction,
described at the fundamental level by quantum chromodynamics (QCD). 
Low-energy phenomena, such as the binding of quarks and gluons within hadrons,
lie in the nonperturbative regime of QCD and are difficult to calculate.
Much progress has been made in recent years in the study of beauty mesons\cite{HFAG}; however, many aspects of beauty baryons are still largely unknown.  Many decays of beauty mesons into pairs of charm hadrons have branching fractions at the percent level~\cite{PDG2012}.  Decays of beauty baryons into pairs of charm hadrons  are expected to be of comparable size, yet none have been observed to date.   If such decays do have sizable branching fractions, they could be used to study beauty-baryon properties.  
For example, a comparison of beauty meson and baryon branching fractions can be used to test factorization in these decays. 

Many models and techniques  have been developed
 that attempt to reproduce the spectrum of the
measured hadron masses,
such as constituent quark models or lattice QCD calculations~\cite{QCD1}.
Precise measurements of ground-state beauty-baryon masses are required to permit 
precision tests of a variety of
QCD models~\cite{QCD2,QCD3,QCD4,QCD5,QCD6,QCD7,QCD8}. 
The \Lb baryon mass is particularly interesting in this context 
since several ground-state beauty-baryon masses are measured relative to that of the \Lb~\cite{LHCb-PAPER-2011-035}.

This Letter reports the first observation of the decays \LbLcDs and \LbLcD made using data corresponding to an integrated luminosity of 
1 and 2\invfb collected at 7 and 8\tev center-of-mass energy in $pp$ collisions, respectively, with the  \lhcb detector. The former is used to make the most precise measurement to date of the mass of the \Lb baryon.
Improved measurements of the branching fraction $\mathcal{B}(\BsDDs)$ and stringent upper limits on $\mathcal{B}(\BdsLcLc)$ are also reported.  Charge conjugated decay modes are implied throughout this Letter.


The \lhcb detector~\cite{Alves:2008zz} is a single-arm forward
spectrometer covering the pseudo\-rapidity
range $2<\eta <5$,
designed for the study of particles containing \bquark or \cquark
quarks. The detector includes a high-precision tracking system
consisting of a silicon-strip vertex detector surrounding the $pp$
interaction region, a large-area silicon-strip detector located
upstream of a dipole magnet with a bending power of about
$4{\rm\,Tm}$, and three stations of silicon-strip detectors and straw
drift tubes~\cite{LHCb-DP-2013-003} placed downstream.
The combined tracking system provides a momentum measurement with
relative uncertainty that varies from 0.4\% at 5\gevc to 0.6\% at 100\gevc,
and impact parameter resolution of 20\mum for
tracks with large transverse momentum (\pt). Different types of charged hadrons are distinguished by information
from two ring-imaging Cherenkov detectors~\cite{LHCb-DP-2012-003}. 
The hardware stage of the trigger uses information from calorimeter and muon systems~\cite{LHCb-DP-2012-004}.  
The calorimeter system consists of scintillating-pad and preshower detectors, an electromagnetic calorimeter and a hadronic calorimeter.  The muon system is composed of alternating layers of iron and multiwire
proportional chambers~\cite{LHCb-DP-2012-002}.
The software stage of the trigger, which applies a full event reconstruction, 
 uses a boosted decision tree (BDT)~\cite{Breiman} to identify secondary vertices consistent with the decay
  of a beauty hadron~\cite{BBDT}.

Samples of simulated events are used to determine the signal selection efficiency, to model signal event distributions and to investigate possible background contributions.
In the simulation, $pp$ collisions are generated using \pythia~\cite{Sjostrand:2006za,*Sjostrand:2007gs} with a specific \lhcb configuration~\cite{LHCb-PROC-2010-056}.  Decays of hadronic particles are described by \evtgen~\cite{Lange:2001uf}, in which final state radiation is generated using \photos~\cite{Golonka:2005pn}.
The interaction of the generated particles with the detector and its response are implemented using the \geant toolkit~\cite{Allison:2006ve, *Agostinelli:2002hh} as described in Ref.~\cite{LHCb-PROC-2011-006}.


In this analysis, signal beauty-hadron candidates are formed by combining charm-hadron candidate pairs reconstructed in the following decay modes: $D^+ \to K^- \pi^+ \pi^+$, $D_s^+ \to K^- K^+ \pi^+$ and $\Lc \to p K^- \pi^+$.  The measured invariant mass of each charm-hadron candidate, the resolution on which is about $6-8$\mevcc, is required to be within 25\mevcc of the nominal value~\cite{PDG2012}.  To improve the resolution of the beauty-hadron mass, the decay chain is fit imposing kinematic and vertex constraints~\cite{Hulsbergen:2005pu}; this includes constraining the charm-hadron masses to their nominal values.   
To suppress contributions from non-charm decays, the
reconstructed charm-hadron decay vertex is required to be downstream
of and significantly displaced from the reconstructed beauty-hadron decay vertex.   

A BDT is used to select each type of charm-hadron candidate.  These BDTs
use five variables for the charm hadron and 23 for
each of its decay products. The variables include kinematic quantities,
track and vertex qualities, and particle identification (PID) information. The
signal and background samples used to train the BDTs
are obtained from  large samples of \BdDpi, \BsDspi and \LbLcpi decays.
These data samples are also used to validate selection efficiencies obtained from simulation.
The signal distributions are background subtracted using weights~\cite{Pivk:2004ty} obtained from
fits to the beauty-hadron invariant mass distributions. The
background distributions are taken from the charm-hadron and high-mass
beauty-hadron sidebands in the same data samples.
To obtain the BDT efficiency in a given signal decay mode, the kinematical
properties and correlations between the two charm hadrons are
taken from simulation.  The BDT response
distributions are obtained from validation data samples of the decays used in the BDT training, weighted to match the kinematics of the signal.

Due to the kinematic similarity of the decays $D^+ \to K^- \pi^+ \pi^+$, $D_s^+ \to K^- K^+ \pi^+$ and $\Lc \to p K^- \pi^+$, cross-feed may occur among beauty-hadron decays into pairs of charm hadrons.
For example, cross-feed between $D^+$ and $D^+_s$ mesons occurs when a $K^- h^+ \pi^+$ candidate is reconstructed in the $D^+$ mass region under the $h^+ = \pi^+$ hypothesis and the $D^+_s$ mass region under the $h^+ = K^+$ hypothesis. 
In such situations, an arbitration is performed:
if the ambiguous track ($h^+$) can be associated to an oppositely-charged track to form a $\phi(1020) \to K^+K^-$ candidate, the kaon hypothesis is taken resulting in a $D_s^+$ assignment to the charm-hadron candidate;
otherwise, stringent PID requirements are applied to $h^+$.  The efficiency of these arbitrations is obtained using simulated signal decays to model the kinematical properties and $D^{*+} \to D^0\pi^+$ calibration data for the PID efficiencies.


Signal yields are determined by performing unbinned extended likelihood fits to the beauty-hadron invariant mass spectra observed in the data.  The signal distributions are modeled using a so-called Apollonios function, which is the exponential of a hyperbola combined with a power-law low-mass tail~\cite{DIEGO}.
The peak position and resolution parameters are allowed to vary while fitting the data, while the low-mass tail parameters are taken from simulation and fixed in the fits.
The measurements reported in this paper are not sensitive to the specific choice of the signal model.   

Four categories of background contributions are considered: 
{\em partially reconstructed} decays of beauty hadrons where at least one final-state particle is not reconstructed; 
 decays into a single charm hadron and three light hadrons; 
 {\em reflections} where the cross-feed arbitration fails to remove a misidentified particle;
and combinatorial background.
The only partially reconstructed decays that contribute in the mass region studied are those where a single pion or photon is not reconstructed; thus, only final states comprised of $D^{*+}_{(s)}$ or ${\it \Sigma}_c^+$ and another charm hadron are considered ({\em e.g.}, $\Lb \to \Lc D^{*-}_s$).  These background contributions are modeled using kernel probability density functions (PDFs)~\cite{KYLE} obtained from simulation.
Single-charm backgrounds are studied using data that are reconstructed outside of a given charm-hadron mass region and are found to be $\mathcal{O}(1\%)$ for decays containing a $D_s^-$ ({\em e.g.}, $\overline{\kern -0.2em B}{}^0 \to D^+ K^-K^+\pi^-$) and negligible otherwise.  
The only non-negligible reflection is found to be \LbLcDs decays misidentified as $\Lc D^-$ candidates.  The shape is obtained from simulation, while the normalization is fixed using simulation and the aforementioned PID calibration sample to determine the fraction of \LbLcDs decays that are not removed by the cross-feed criteria.
Reflections of \BdDDs decays misidentified as final states containing $\Lc$ particles do not have a peaking structure and, therefore, are absorbed into the combinatorial backgrounds, which are modeled using exponential distributions.

\begin{figure}[t]
\centering
\includegraphics[width=0.49\textwidth]{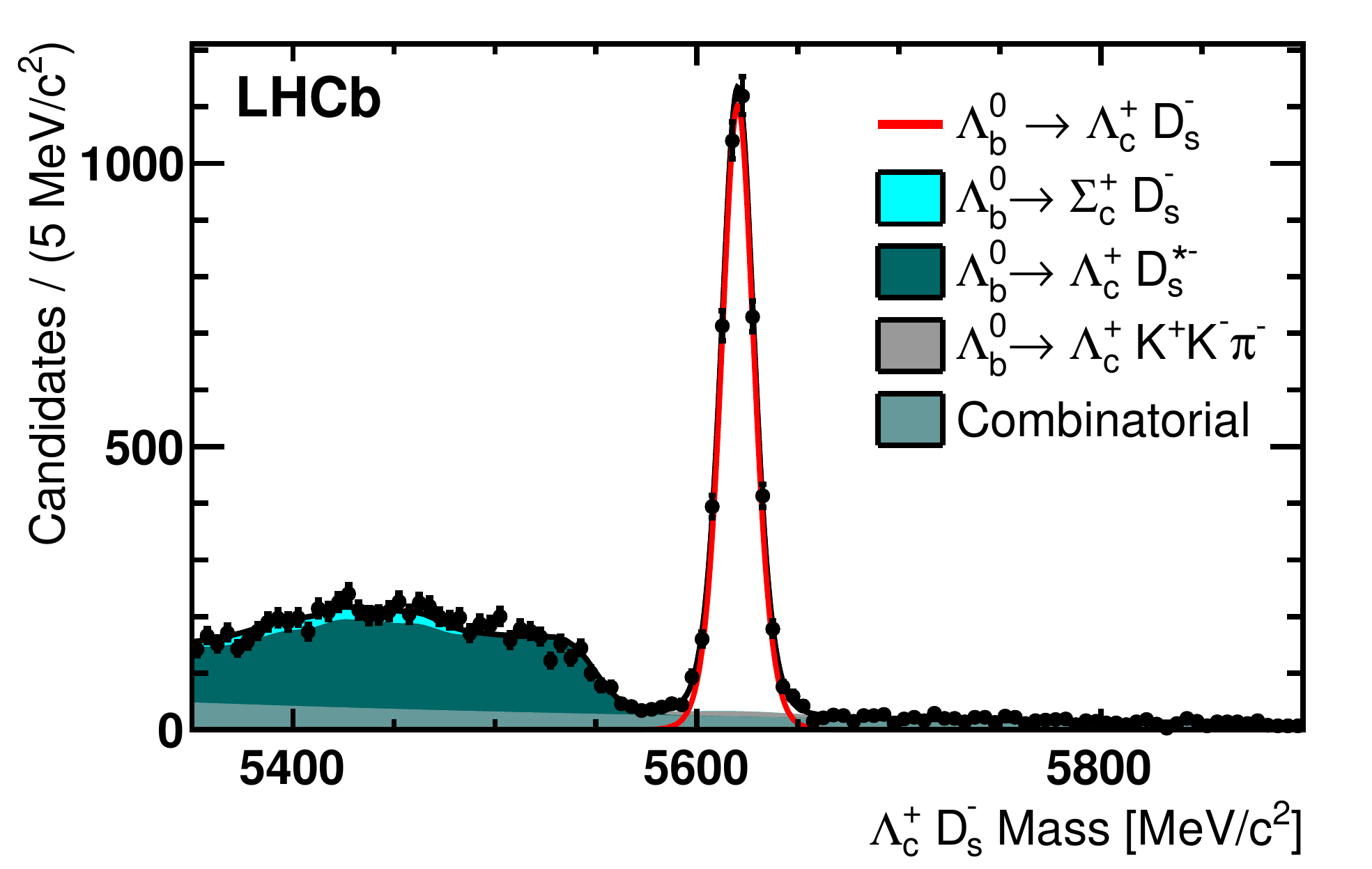}
\includegraphics[width=0.49\textwidth]{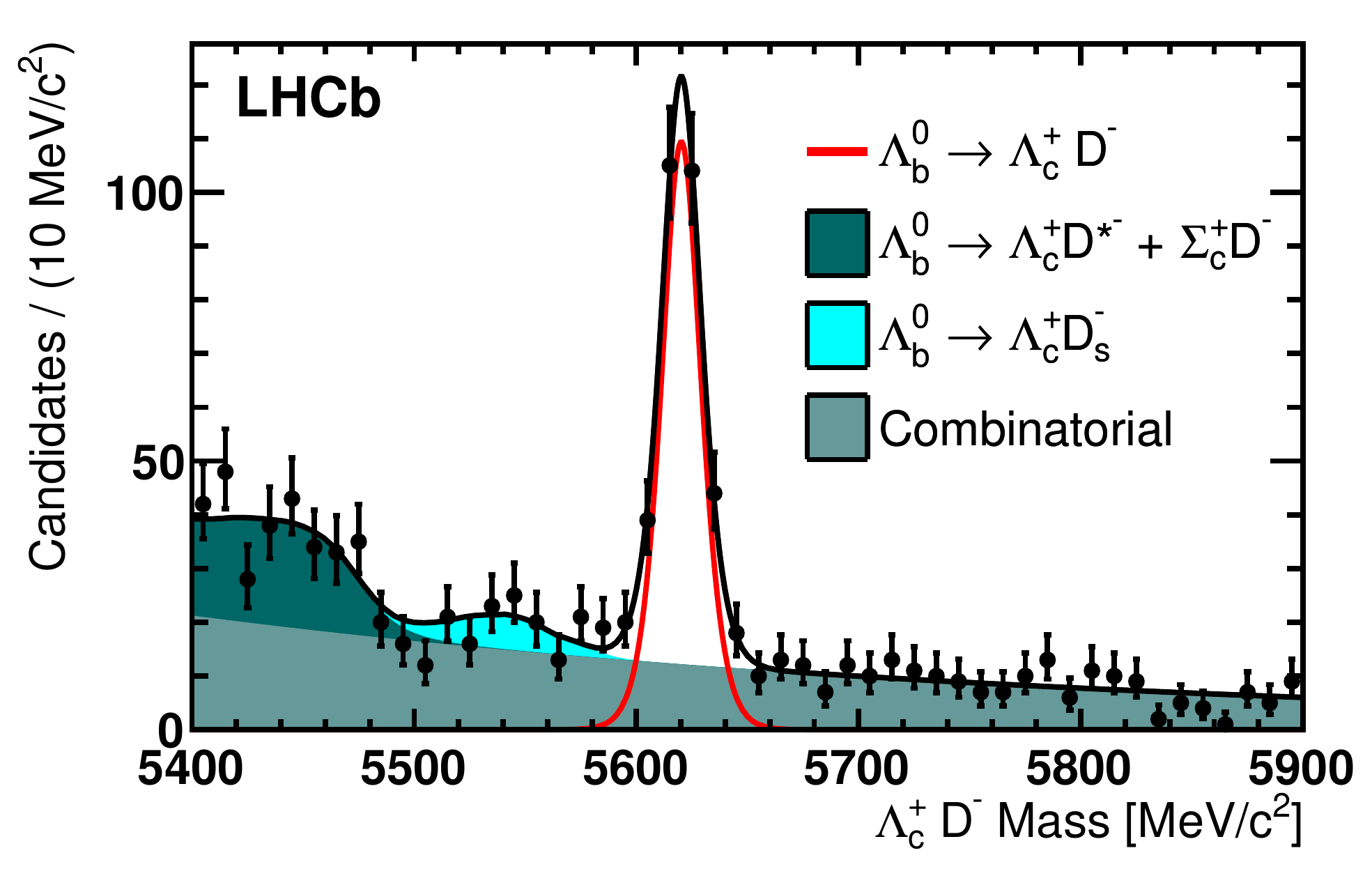}
\caption{
\label{fig:lcds}
Invariant mass distributions for (left) \LbLcDs and (right) \LbLcD candidates with the fits described in the text overlaid.  
}
\end{figure}

Figure~\ref{fig:lcds} shows the invariant mass spectra for the \LbLcDs and \LbLcD candidates.  
The signal yields obtained are $4633\pm69$ and $262\pm19$ for \LbLcDs and \LbLcD, respectively.  This is the first observation of each of these decays.
The ratio of branching fractions determined using the nominal $D_s^-$~\cite{PDG2012} and $D^-$~\cite{DBR} meson branching fractions and the ratio of efficiencies is 
\begin{equation}
\frac{\mathcal{B}(\LbLcD)}{\mathcal{B}(\LbLcDs)} = 0.042 \pm 0.003\stat \pm 0.003\syst. \nonumber 
\end{equation}
The similarity of the final states and the shared parent particle result in many cancelations of uncertainties in the determination of the ratio of branching fractions.
The remaining uncertainties include roughly equivalent contributions from determining the efficiency-corrected yields and from the ratio of charm-hadron branching fractions (see Table~\ref{tab:bf_sys}).  The dominant contribution to the uncertainty of the fit PDF is due to the low-mass background contributions, which are varied in size and shape to determine the effect on the signal yield.  
The efficiencies of the cross-feed and BDT criteria are determined in a data-driven manner that produces small uncertainties.  
The observed ratio is approximately the ratio of the relevant quark-mixing factors and meson decay constants,
$|V_{cd}/V_{cs}|^2 \times (f_D/f_{D_s})^2 \approx 0.034$,
as expected assuming nonfactorizable effects are small.

\begin{table}[t]
  \caption{Relative systematic uncertainties on branching fraction measurements (\%). 
The production ratio $\sigma(\Bs)/\sigma(\Bd)$ is taken from Ref.~\cite{LHCb-CONF-2013-011}.
Numbers in brackets in the last column are for the $B^0_s$ decay mode. 
} 
\begin{center}
\vspace{-0.1in}
\begin{tabular}{ccccc}
    \hline \hline
Source & $\frac{\mathcal{B}(\LbLcD)}{\mathcal{B}(\LbLcDs)}$ & \hspace{-0.05in}$\left[\frac{\mathcal{B}(\LbLcDs)}{\mathcal{B}(\BdDDs)}\right]\hspace{-0.01in}\big/\hspace{-0.01in}\left[\frac{\mathcal{B}(\LbLcpi)}{\mathcal{B}(\BdDpi)}\right]$\hspace{-0.05in} & $\frac{\mathcal{B}(\BsDDs)}{\mathcal{B}(\BdDDs)}$ & $\frac{\mathcal{B}(\BdsLcLc)}{\mathcal{B}(\BdsDDs)}$ \\
\hline
Efficiency & 3.5 & 5.2 & 1.0 & 3.9 (5.0) \\
Fit model & 3.0 & 2.6 & 3.0 & $-$ \\
$\mathcal{B}(D^+_{(s)},\Lc)$ & 5.2 & $-$ & $-$ & 8.8 \\
$\sigma(\Bs)/\sigma(\Bd)$ & $-$ & $-$ & 5.8 & $-$ \\
\hline
Total & 6.9 & 5.8 & 6.6 & 9.6 (10.1) \\
\hline \hline
\end{tabular}\end{center}
\label{tab:bf_sys}
\end{table}

\begin{figure}[t!]
\centering
\includegraphics[width=0.49\textwidth]{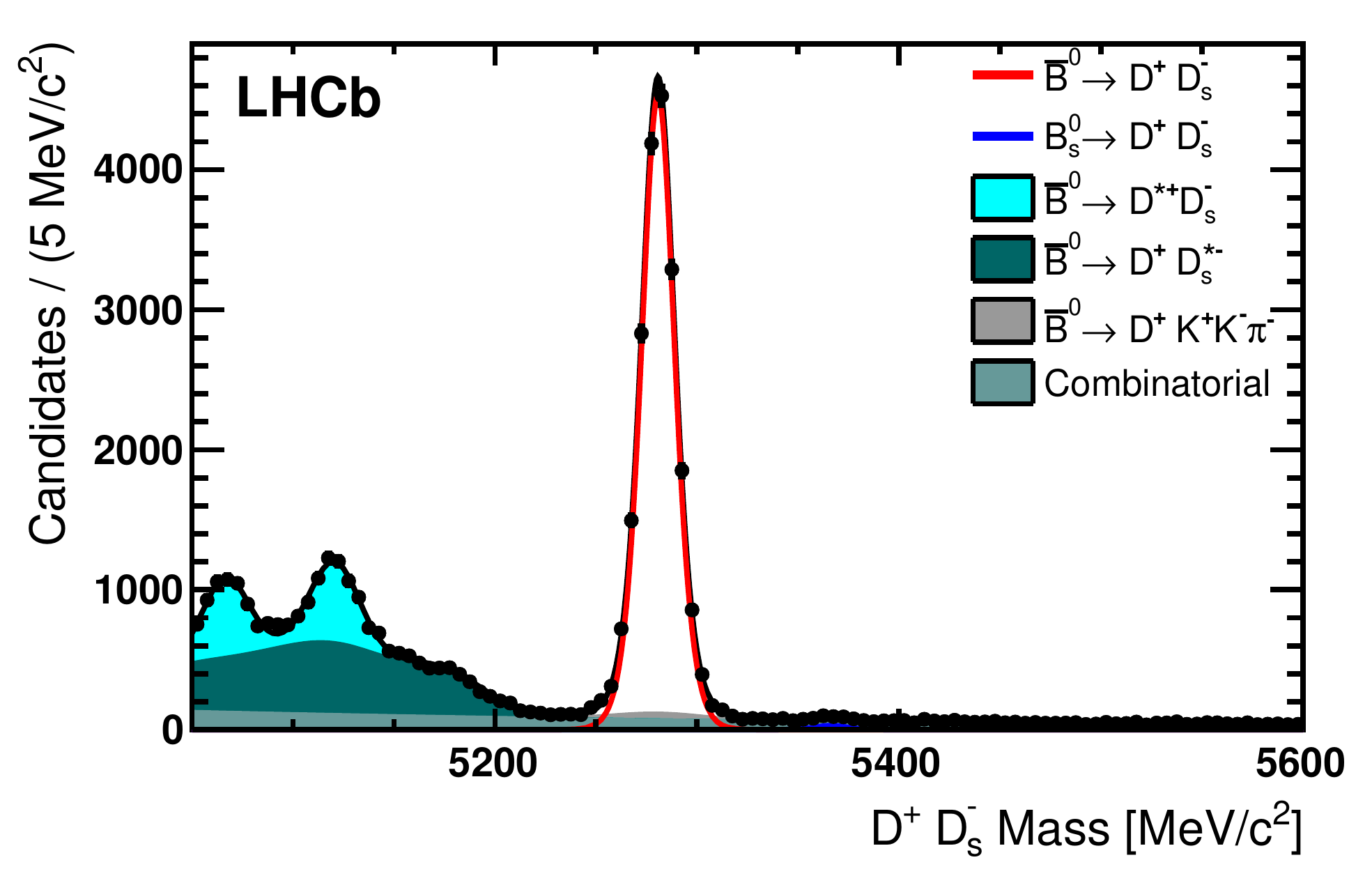}
\includegraphics[width=0.49\textwidth]{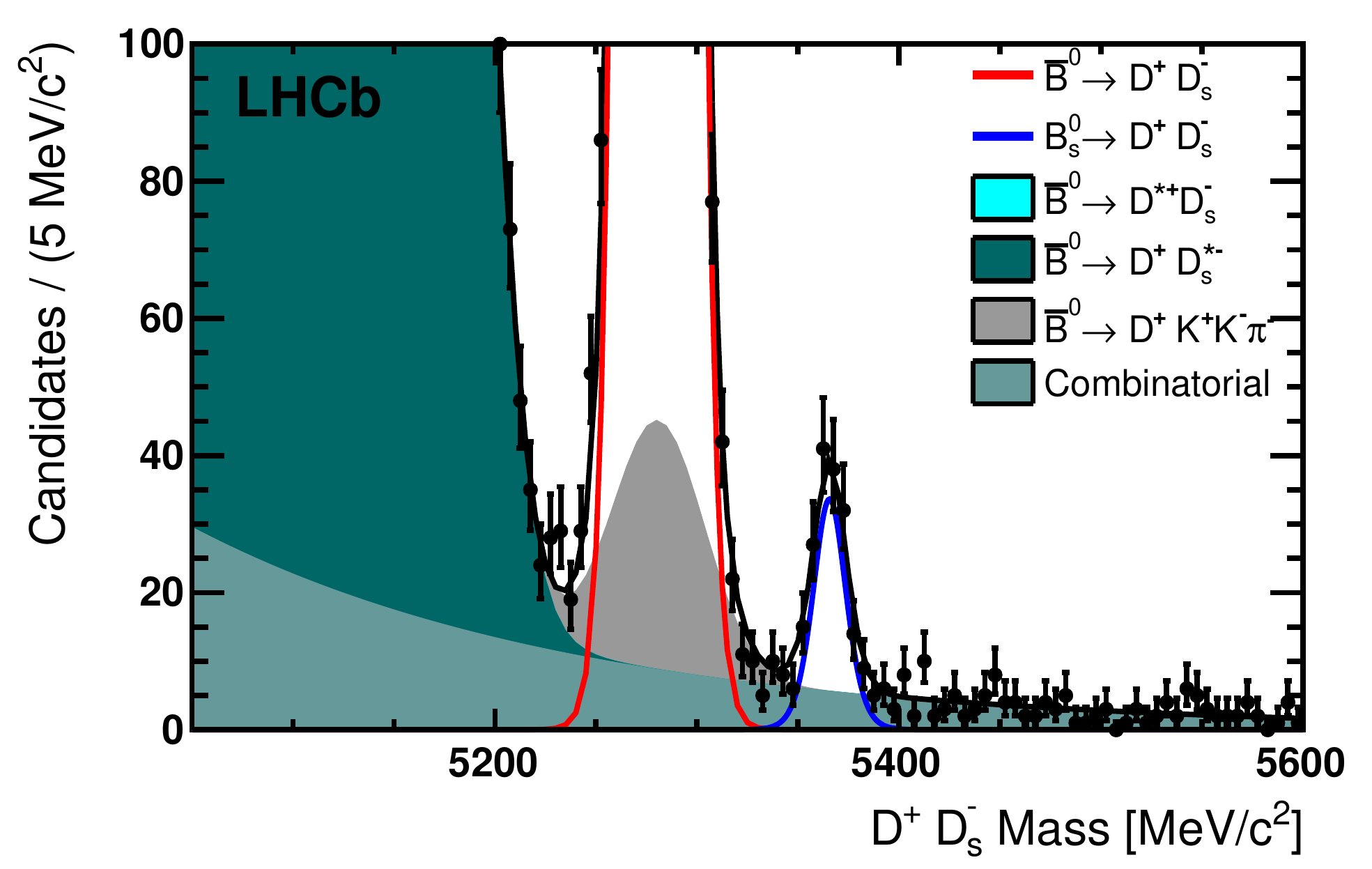}
\caption{
\label{fig:dds}
Invariant mass distributions for $D^+D_s^-$ candidates selected using BDT criteria optimized for the (left) \BdDDs and (right) \BsDDs decay modes with the fits described in the text overlaid.
}
\end{figure}

\begin{figure}[t!]
\centering
\includegraphics[width=0.49\textwidth]{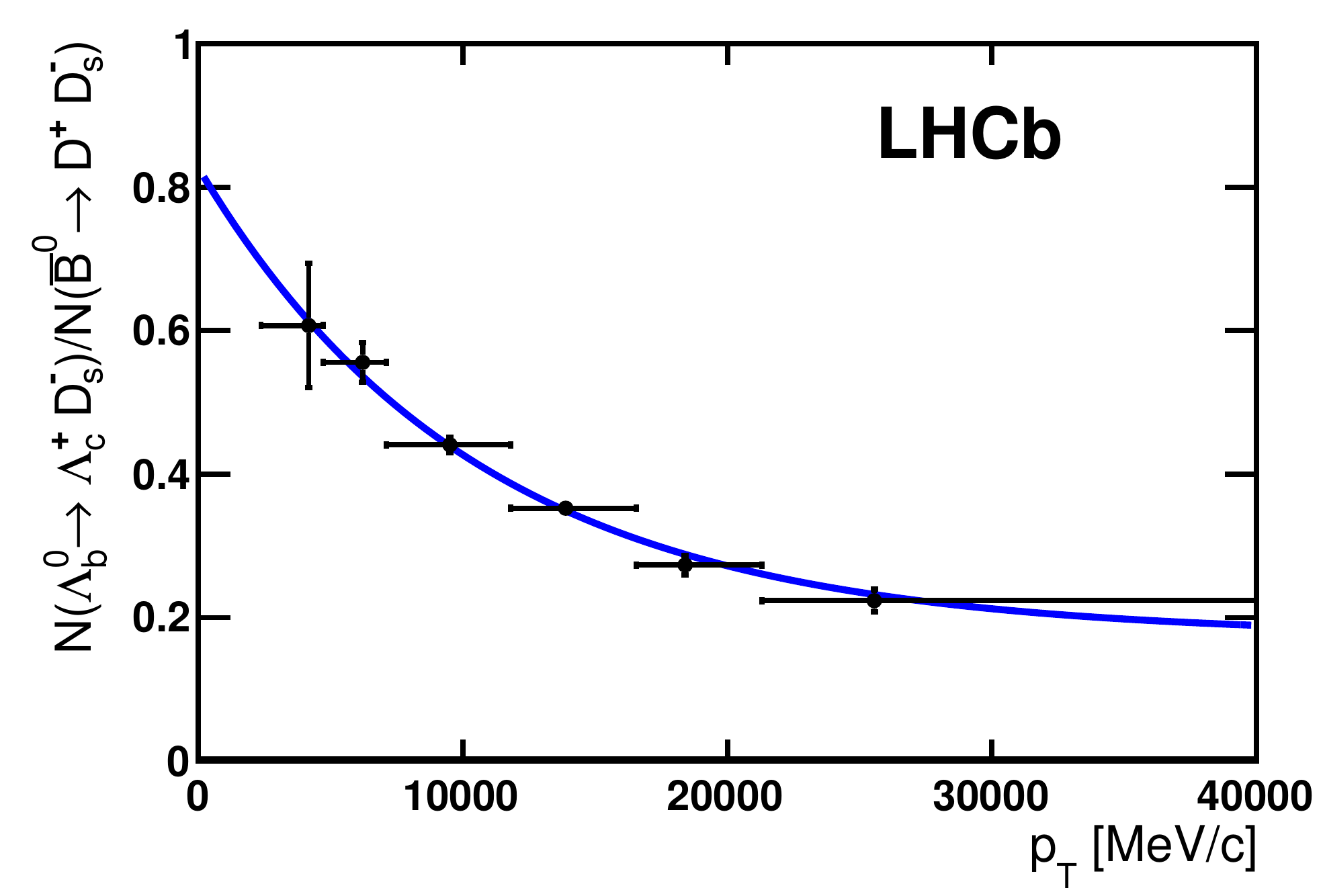}
\caption{
\label{fig:lcds_dsd}
Efficiency-corrected yield ratio of $\LbLcDs$ to $\BdDDs$ vs \pt.  The points are located at the mean \pt value of the \Lb in each bin.  
The curve shows the data fit with the shape of the \pt dependence
measured in Ref.~\cite{LCPI}.
}
\end{figure}

The branching fraction of the decay \LbLcDs is determined relative to that of the \BdDDs decay.  Using $D^+ D_s^-$ BDT criteria, optimized to maximize the expected \Bd significance, $19\,395\pm145$ \BdDDs decays are observed  (see Fig.~\ref{fig:dds}).
The measurement of $\mathcal{B}(\LbLcDs)/\mathcal{B}(\BdDDs)$ is complicated by the fact that the ratio of the \Lb and \Bd production cross sections, $\sigma(\Lb)/\sigma(\Bd)$, depends on the \pt of the beauty hadrons\cite{LHCb-PAPER-2011-018}.  Figure~\ref{fig:lcds_dsd} shows the ratio of efficiency-corrected yields, $N(\LbLcDs)/N(\BdDDs)$, as a function of beauty-hadron \pt. 
The ratio of branching-fraction ratios is obtained  using a fit with the shape of the \pt dependence measured in $\mathcal{B}(\LbLcpi)/\mathcal{B}(\BdDpi)$~\cite{LCPI} and found to be
\begin{equation}
\left[\frac{\mathcal{B}(\LbLcDs)}{\mathcal{B}(\BdDDs)}\right]\big/\left[\frac{\mathcal{B}(\LbLcpi)}{\mathcal{B}(\BdDpi)}\right] = 0.96 \pm 0.02\stat \pm 0.06\syst. \nonumber
\end{equation}
This result does not depend on the absolute ratio of production cross sections or on any charm-hadron branching fractions.  The systematic uncertainties on this result are listed in Table~\ref{tab:bf_sys}.  The uncertainty in the fit model is largely due to the sizable single-charm contributions to these modes and due to contributions from the fits described in Ref.~\cite{LCPI}.  
The ratio $N(\LbLcDs)/N(\BdDDs)$ is observed to be consistent in data collected at $\sqrt{s} = 7$ and 8\tev; thus, it is assumed that the production fractions of the \Lb and \Bd are the same for all data analyzed and no systematic uncertainty is assigned.
The ratio of branching ratios is consistent with unity as expected assuming small nonfactorizable effects.

The kinematic similarity of the decay modes \LbLcDs and \BdDDs permits a precision measurement of the mass difference of the \Lb and \Bd hadrons to be made.  The relatively small value of $\left[M(\Lb)-M(\Lc)-M(D_s^-)\right]-\left[M(\Bd)-M(D^+)-M(D_s^-)\right]$ means that the uncertainty due to momentum scale, the dominant uncertainty in absolute mass measurements, mostly cancels; however, it is still important to determine accurately the momenta of the final state particles.  The momentum scale calibration of the spectrometer, which accounts for imperfect knowledge of the magnetic field and alignment, is discussed in detail in Refs.~\cite{LHCb-PAPER-2011-035,LHCb-PAPER-2013-011}.  The uncertainty on the calibrated momentum scale is estimated to be 0.03\% by comparing various particle masses measured at \lhcb to their nominal values~\cite{LHCb-PAPER-2013-011}.  

The kinematic and vertex constraints used in the fits described previously reduce the statistical uncertainty on $M(\Lb) - M(\Bd)$ by improving the resolution.  These constraints also increase the systematic uncertainty by introducing a dependence on the precision of the nominal charm-hadron masses.  For this reason, these constraints are not imposed in the mass measurement.   The mass difference obtained is 
\begin{equation}
  M(\Lb) - M(\Bd) = 339.72 \pm 0.24\stat \pm 0.18\syst\mevcc. \nonumber
\end{equation}
The dominant systematic uncertainty (see Table~\ref{tab:mass_sys}) arises due to a correlation between the reconstructed beauty-hadron mass and charm-hadron flight distance.  
The large difference in the \Lc and $D^+$ hadron lifetimes\cite{PDG2012} causes only a partial cancelation of the biases induced by the charm-lifetime selection criteria.
This effect is studied in simulation and a 0.16\mevcc uncertainty is assigned.
The 0.03\% uncertainty in the momentum scale results in an 
uncertainty on the mass difference of
0.08\mevcc.  
Many variations in the fit model are considered and none produce a significant shift in the mass difference.
The systematic uncertainty in the mass difference due to the uncertainty in the amount of detector material in which charged particles lose energy is negligible~\cite{LHCb-PAPER-2013-011}.
Furthermore,  the uncertainty on $M(\Lb) - M(\Bd)$ due to differences in beauty-hadron production kinematics, as seen in Fig.~\ref{fig:lcds_dsd}, is also found to be negligible.  

Using the nominal value for $M(\Bd)$\cite{PDG2012} gives $M(\Lb) = 5619.30\pm0.34$\mevcc, where the uncertainty includes both statistical and systematic contributions.  This is the most precise result to date.  
The total uncertainty is dominated by statistics and charm-hadron lifetime effects; thus, this result can be treated as being uncorrelated with the previous \lhcb result obtained using the $\Lb \to J/\psi {\it \Lambda}^0$ decay~\cite{LHCb-PAPER-2012-048}.  A weighted average of the \lhcb results gives $M(\Lb) = 5619.36\pm0.26$\mevcc.  This value may then be used to improve the precision of the ${\it \Xi}_b^-$ and ${\it \Omega}_b^-$ baryon masses using their mass differences with respect to the \Lb baryon as reported in Ref.~\cite{LHCb-PAPER-2012-048}.

\begin{table}
  \caption{\label{tab:mass_sys} Systematic uncertainties for $M(\Lb) - M(\Bd)$. 
  }
\begin{center}
\vspace{-0.1in}
\begin{tabular}{cc}
\hline    \hline  
Description & Value (\mevcc) \\
\hline
$\Lc-D^+$ lifetime difference & 0.16\\
Momentum scale & 0.08 \\
Fit model & 0.02 \\
\hline  
Total & 0.18 \\
\hline \hline
\end{tabular}
\end{center}
\end{table}

Using BDT criteria optimized for maximizing the expected significance of ${\BsDDs}$, $14\,608\pm121$ \Bd and $143\pm14$ $B_s^0$ decays are observed (see Fig.~\ref{fig:dds}), from which the ratio extracted is
\begin{equation}
\frac{\mathcal{B}(\BsDDs)}{\mathcal{B}(\BdDDs)} = 0.038\pm0.004\stat\pm0.003\syst. \nonumber
\end{equation}
This yields the most precise measurement to date of $\mathcal{B}(\BsDDs)$ and supersedes Ref.~\cite{LHCb-PAPER-2012-050}.  
The systematic uncertainty is dominated by the contribution from beauty-hadron production fractions since the two decay modes share the same final state.  The additional small efficiency uncertainty is due to the uncertainty on the $B_s^0$ lifetime.  The fit model uncertainty is largely due to the size of the combinatorial background near the $B_s^0$ peak. 
The ratio of branching fractions is approximately the ratio of quark-mixing factors as expected assuming nonfactorizable effects are small.

A search is also performed for the decay modes \BdsLcLc.  Regions centered around the nominal beauty-meson masses with boundaries defined such that each region contains 95\% of  the corresponding signal are determined using simulation.  The expected background contribution in each of these regions is obtained from the charm-hadron mass sidebands.   
Applying this technique to the \BdDDs and $\Lb \to \Lc D^-_{(s)}$ decays produces background estimates consistent with those obtained by fitting the invariant mass spectra for those modes.  
 No significant excess is observed in either $\Lc{\it \Lambda}_c^-$ signal region.  The limits obtained using the method of Ref.~\cite{Rolke} and the known $D_s^-$~\cite{PDG2012}, $D^-$~\cite{DBR} and $\Lc$~\cite{LCBR} hadron branching fractions are
\begin{eqnarray}
\frac{\mathcal{B}(\BdLcLc)}{\mathcal{B}(\BdDDs)} & < & 0.0022 \; [95\% \; {\rm C.L.}], \nonumber\\
\frac{\mathcal{B}(\BsLcLc)}{\mathcal{B}(\BsDDs)} & < & 0.30 \;  [95\% \; {\rm C.L.}]. \nonumber
\end{eqnarray}
For these results the lifetime of the light-mass \Bs eigenstate is assumed as this produces the most conservative limits\cite{HFAG}. 
This is the best limit to date for the \Bd decay mode and the first limit for the $B_s^0$ decay mode. 


In summary, first observations and relative branching fraction measurements have been made for the decays $\Lb \to \Lc D^-_{(s)}$.  
The most precise measurement of the \Lb baryon mass has been made via its mass difference relative to the \Bd meson. 
The most precise measurement of $\mathcal{B}(\BsDDs)$ has been presented and the most stringent upper limits have been placed on $\mathcal{B}(\BdsLcLc)$.  Using the PDG value $\mathcal{B}(\BdDDs) = (7.2\pm0.8)\times 10^{-3}$~\cite{PDG2012} and the \lhcb result for ${\mathcal{B}(\LbLcpi)}/{\mathcal{B}(\BdDpi)}$~\cite{LCPI}, the absolute branching fractions obtained are
\begin{eqnarray}
\mathcal{B}(\LbLcDs) &=& (1.1 \pm 0.1) \times 10^{-2}, \nonumber \\
\mathcal{B}(\LbLcD) &=& (4.7 \pm 0.6) \times 10^{-4}, \nonumber \\
\mathcal{B}(\BsDDs) &=& (2.7 \pm 0.5) \times 10^{-4},  \nonumber\\
\mathcal{B}(\BdLcLc) &<& 1.6 \times 10^{-5} \, [95\% \, {\rm C.L.}],  \nonumber\\
\mathcal{B}(\BsLcLc) &<& 8.0 \times 10^{-5} \, [95\%\, {\rm C.L.}].  \nonumber 
\end{eqnarray}
These results are all consistent with expectations that assume small nonfactorizable effects. 

\section*{Acknowledgements}

\noindent We express our gratitude to our colleagues in the CERN
accelerator departments for the excellent performance of the LHC. We
thank the technical and administrative staff at the LHCb
institutes. We acknowledge support from CERN and from the national
agencies: CAPES, CNPq, FAPERJ and FINEP (Brazil); NSFC (China);
CNRS/IN2P3 and Region Auvergne (France); BMBF, DFG, HGF and MPG
(Germany); SFI (Ireland); INFN (Italy); FOM and NWO (The Netherlands);
SCSR (Poland); MEN/IFA (Romania); MinES, Rosatom, RFBR and NRC
``Kurchatov Institute'' (Russia); MinECo, XuntaGal and GENCAT (Spain);
SNSF and SER (Switzerland); NAS Ukraine (Ukraine); STFC (United
Kingdom); NSF (USA). We also acknowledge the support received from EPLANET and the
ERC under FP7. The Tier1 computing centres are supported by IN2P3
(France), KIT and BMBF (Germany), INFN (Italy), NWO and SURF (The
Netherlands), PIC (Spain), GridPP (United Kingdom).
We are indebted to the communities behind the multiple open source software packages we depend on.
We are also thankful for the computing resources and the access to software R\&D tools provided by Yandex LLC (Russia).

\addcontentsline{toc}{section}{References}
\setboolean{inbibliography}{true}

\ifx\mcitethebibliography\mciteundefinedmacro
\PackageError{LHCb.bst}{mciteplus.sty has not been loaded}
{This bibstyle requires the use of the mciteplus package.}\fi
\providecommand{\href}[2]{#2}

\end{document}